\documentclass[12pt]{spieman}  
\usepackage{amsmath,amsfonts,amssymb}
\usepackage{graphicx}
\usepackage{setspace}
\usepackage{tocloft}

\usepackage[sort&compress,square,comma,numbers]{natbib}

\DeclareRobustCommand{\citeext}[1]{\cite{#1}}

\usepackage{circuitikz}
\usepackage{tikz}
\usetikzlibrary{decorations.markings}
\newif\iflabrev

\usepackage{amsmath}
\usepackage{amssymb}
\usepackage{mathtools}
\usepackage{bigints}

\title{Suitability of Magnetic Microbolometers based on Paramagnetic Temperature Sensors for CMB Polarization Measurements}

\author[a,b,d,*]{J. M. Geria}
\author[a,d]{M. R. Hampel}
\author[b,c]{S. Kempf}
\author[e]{J. Bonaparte}
\author[a,c,d]{L. P. Ferreyro}
\author[a,c,d]{M. Garcia Redondo}
\author[a,d]{A. Almela}
\author[a,c,d]{J. M. Salum}
\author[a,b,e]{N. Müller}
\author[a,b]{J. Bonilla-Neira}
\author[a,d]{A. Fuster}
\author[a]{M. Platino}
\author[a]{A. Etchegoyen}
\affil[a]{Instituto de Tecnologías en Detección y Astropartículas (CNEA, CONICET, UNSAM), Buenos Aires, Argentina}
\affil[b]{Institute of Micro- and Nanoelectronic Systems (IMS), Karlsruhe Institute of Technology (KIT), Karlsruhe, Germany}
\affil[c]{Institute for Data Processing and Electronics (IPE), Karlsruhe Institute of Technology (KIT), Karlsruhe, Germany}
\affil[d]{Universidad Tecnológica Nacional, Facultad Regional Buenos Aires, Argentina}
\affil[e]{Comisión Nacional de Energía Atómica (CNEA), Buenos Aires, Argentina}

\cftpagenumbersoff{figure}
\cftpagenumbersoff{table} 
\begin{document} 
\maketitle

\begin{abstract}
High-resolution maps of polarization anisotropies of the cosmic microwave background (CMB) are in high demand, since the discovery of primordial B-modes in the polarization patterns would confirm the inflationary phase of the universe that would have taken place
before the last scattering of the CMB at the recombination epoch. Transition edge sensors (TES) and microwave kinetic inductance detectors (MKID) are the predominant detector technologies of cryogenic detector array-based CMB instruments that search for primordial B-modes. We propose another type of cryogenic detector to be used for
CMB survey: a magnetic microbolometer (MMB) that is based on a paramagnetic temperature sensor. It is an adaption of state-of-the-art metallic magnetic calorimeters (MMCs) that are meanwhile a key technology for high resolution $\alpha$, $\beta$, $\gamma$ and X-ray spectroscopy as well as the study of neutrino mass. The effort to adapt MMCs for CMB surveys is triggered by their lack of Johnson noise associated with the detector readout, the possibility of straightforward calibration and higher dynamic range given it possesses a broad and smooth responsivity dependence with temperature and the absence of Joule dissipation which simplifies the thermal design.
A brief proof of concept case study is analyzed, taking into account typical constraints in CMB measurements and reliable microfabrication processes, to assess the suitability of metallic magnetic sensors in CMB experiments. The results show that MMBs provide a promising technology for CMB polarization survey as their sensitivity can be tuned for background limited detection of the sky while simultaneously maintaining a low time response to avoid distortion of the point-source response of the telescope. As the sensor technology and its fabrication techniques are compatible with TES based bolometric detector arrays, a change of detector technology would even come with very low cost.

\end{abstract}


{\noindent \footnotesize\textbf{*}Email,  \linkable{juan.geria@iteda.cnea.gov.ar} }

\begin{spacing}{1}   

\section{Introduction}

Ever since the first cosmic microwave background (CMB) measurements around 60 years ago were taken, much has been learned. The data point to a big bang origin of the universe. The discovery of temperature anisotropies hints to late-time large structure evolution and primordial quantum fluctuations. The scale of the anisotropies also tells us about the composition of the universe and its geometry; however, open questions still remain. One of the most relevant questions nowadays is whether CMB polarization encodes an imprint of primordial gravitational waves, which would be the conclusive evidence of cosmic inflation.

Polarization patterns imprinted on the CMB can be decomposed into $E$- and $B$-modes components, primordial $B$-modes being the unique direct observational signature of the inflationary phase \cite{Spergel1997, Seljak1997} that is thought to have taken place in the early Universe, generating primeval perturbations, producing elementary particles predicted by the standard model of particle physics and giving the Universe its generic features. Various instruments are being developed to measure the polarization anisotropies, among these are QUBIC \cite{Battistelli2011b}, SPTPol \cite{Keisler2015}, POLARBEAR \cite{Nishino2012}, BICEP2 \cite{Ade2014}, CLASS \cite{Essinger-Hileman2014}, POLARBEAR 2 + Simons Array \cite{Suzuki2016}, advanced ACTPol \cite{Henderson2016} and many others.

Cryogenic bolometers are thermal radiation detectors that can be designed to measure power flux of electromagnetic waves over a wide range of wavelengths. They turn out to be particularly well suited for sub-millimeter and millimeter-wave applications, where they are presently the most sensitive broadband detectors available. These features combined with the maturity of micro- and nanofabrication technology allowing to produce thousands of virtually identical detectors as well as sophisticated readout techniques, makes them ideal candidates for realizing large-scale detector arrays requested by modern radio telescopes. The unmatched sensitivity over a wide spectral range of cryogenic bolometers will allow to characterize the polarization anisotropies of the CMB with the accuracy needed in the search for primordial B-modes.

Transition-Edge Sensors (TES) are the workhorse of state-of-the-art CMB detector arrays now \cite{Goldie2008, Audley2004}. A TES is a resistive temperature sensor that consists of a superconducting structure made from a single or multilayered thin film superconducting material biased at its normal to superconducting state transition. This transition presents a sudden and abrupt drop in resistance to zero; therefore, small changes in temperature can be measured as large resistance variations.

Even though these detectors are continuously used not only in CMB observations but various other applications, they do still present challenges when implemented in large detector arrays.  The transition temperature, $T_{\text{c}}$, and critical current, $I_{\text{c}}$, of superconducting multilayers become difficult to control even in modern microfabrication facilities. Even differences in the order of nanometers in thickness can produce noticeable $T_{\text{c}}$ and $I_{\text{c}}$ shifts. Large cryogenic detector arrays demand readout electronics with large multiplexing factors to minimize heat load, when noticeable variations in the detectors throughout the array are present, readout and biasing circuits need to be calibrated against position dependent characteristics. These calibration procedures are quite complex and need large amounts of calibration data, in addition to the reported lack of reproducibility of TESs between cooldowns \cite{Hoteling2009}. Furthermore, TES are also known to count with limited dynamic range, BICEP2 and POLARBEAR, for instance, combined two TES sensors in each pixel with a different $T_{\text{c}}$ and saturation power adjusted for different dynamic ranges to overcome this issue \cite{Ade2015, Arnold2012}.  

An alternative to TESs is Microwave Kinetic Inductance Detectors (MKIDs) \cite{Lee2020, Dibert2022}. They are based on the variation of the kinetic inductance of a superconducting strip or coil due to the breaking of Cooper Pairs after absorption of radiation. Incoming radiation produces a change in the inductance and when used, for example, as part of a microresonator the measured resonance frequency shifts can be interpreted as absorbed incoming signal. MKIDs are particularly attractive as they do not need to be biased with a constant voltage or current like TES or need multiple signal calibration procedures. 

Magnetic microcalorimeters (MMCs) based on paramagnetic temperature sensors are an alternative detector technology that became prominent in recent years, especially in the field of high-resolution X-ray spectroscopy and the investigation of the neutrino mass \cite{Fleischmann2005, Fleischmann2009, Kempf2018}. They are cryogenic single particle detectors that are extensively used for measuring the energy of ionizing radiation with very high precision. They are composed of an absorber suited for the particles to be detected that is in tight thermal contact to a paramagnetic temperature sensor. The latter is weakly coupled to a thermal bath and situated in a weak external magnetic field to create a temperature-dependent sensor magnetization. The tiny change of sensor magnetization upon the absorption of an energetic particle is sensed as change of magnetic flux by using a superconducting flux transformer that is inductively coupled to the sensor and a low noise superconducting quantum interference device (SQUID). This unique combination of a paramagnetic thermometer and a near-quantum limited amplifier results in a calorimeter with excellent energy resolution, fast intrinsic signal rise time, almost ideal linear detector response, quantum efficiency close to 100\% as well as a huge dynamic range \cite{Kempf2018}. However, though MMCs are routinely used for calorimetric applications, to present date, they have never been used to tackle bolometric applications.

Against this background, we study whether bolometers based on paramagnetic temperature sensors which we call hereinafter magnetic microbolometers (MMBs) have sufficient sensitivity for bolometric applications and are hence suitable alternatives for CMB polarization measurements. In this context, we particularly take into account that the MMBs discussed within this paper are ultimately intended for installation in cryostats of on-going or planned CMB instruments that are operated at radiotelescopes and usually reach a base temperature in the range of approximately $100\,\text{mK} - 350\,\text{mK}$. This is significantly higher than the operation temperature of state-of-the-art MMCs that is about $10\,\text{mK} - 50\,\text{mK}$ \cite{Kempf2018}. For the comprehensive study of this detector concept, we first summarize the theoretical framework that we have used to forecast the sensor performance and discuss afterwards whether an MMB based detector can achieve background limited detection.

\section{Bolometric Detection in CMB Observations}

A bolometer consists of an absorber or thermal mass that absorbs incident radiation power $P$ and thermalizes the energy, a weak thermal link to a heat bath that keeps the absorber temperature at some defined value in the absence of a power input, and a thermometer in tight thermal contact to the absorber that measures its temperature increase $\Delta T$, which may be determined as\cite{Richards1994}
\begin{equation}
    \Delta T = \frac{P}{G_{\text{bath}}},
    \label{eqRespBasic}
\end{equation}
where $G_{\text{bath}}$ is the heat conductance of the thermal link.

The time constant of such a detector, i.e. the time the detector takes to return to heat bath temperature switching of the constant power input, sets the speed of the bolometer and is given as
\begin{equation}
    \tau = \frac{C_{\text{abs}}}{G_{\text{bath}}},
    \label{eqTimeBasic}
\end{equation}

where $C_{abs}$ is the heat capacity of the absorber or thermal mass. It can be noted from Eq.\,\ref{eqRespBasic} and \ref{eqTimeBasic} that decreasing the value of $G_{\text{bath}}$ increases the thermal gain at the expense of increasing the time constant. There exists hence a trade-off between gain and speed when optimizing a cryogenic bolometer. The optimization process starts by first achieving a weak thermal coupling to produce a high thermal gain, and later reducing the heat capacity of the thermal mass to achieve the required detection speed.

Constraints for CMB detectors impose two main goals, i.e. high sensitivity and low response time. High sensitivity is required to achieve background limited photometry (BLIP) and a fast response to avoid distortion of the point-source response of the telescope. CMB observations are designed with a specific observation strategy in mind. Observation strategies, mechanical limitations of the instrument, required angular scale resolution and avoidance of 1/f noise, typically present in this type of observation, set the needed scan speed $\Dot{\Theta}$ (in deg/sec) at which the telescope must be slewed throughout the portion of the sky that is observed. To avoid the mentioned distortion and in order to respect the Nyquist-Shannon theorem, it can be demonstrated that the time constant $\tau$ of the detector must be \cite{Hanany1998}:

\begin{equation}
    \tau \leq \frac{\Theta_{\text{beam}}}{2\pi \Dot{\Theta}},
    \label{eqMaxTimeConstant}
\end{equation}
where $\Theta_{\text{beam}}$ is the full width at half maximum (FWHM) in degrees of the detector's beam intensity.

A figure of merit commonly used to characterize CMB detectors is the noise equivalent power (NEP) defined as the minimum required power spectral density at the input of the detector to achieve a signal-to-noise ratio (SNR) of one in one Hz of output bandwidth.

Noise in bolometric measurements is composed of the random fluctuation in the arrival of photons to the detector referred to as photon or background noise $\text{NEP}_{\gamma}$ and the intrinsic noise $\text{NEP}_{\text{det}}$ of the detector itself. It is desirable to reduce the detector noise $\text{NEP}_{\text{det}}$ as much as possible until $\text{NEP}_{\gamma}$ is the dominant noise contribution. At this point, further reduction of the $\text{NEP}_{\text{det}}$ does not account for better sensitivity of an individual detector and under this condition the detector is said to be background-limited. The sensitivity of a background-limited instrument can only be improved by increasing the integration time, either by planning longer experiments or increasing the number of detectors that measure simultaneously in the array. This is why, for CMB instruments, it is generally required to have both background-limited and scalable detectors.

To estimate the $\text{NEP}_{\gamma}$, one must first know the average incoming optical power $P_{\text{opt}}$. This is calculated accounting not only for the source radiation (in this case the CMB), but every element that lies between the source and the detector. All optical components and atmosphere (for ground-based experiments) will present a contribution to $P_{\text{opt}}$ correspondent to their emissivity, optical efficiency and spillover losses. Once $P_{\text{opt}}$ is known, assuming a single-mode detector, we can estimate $\text{NEP}_{\gamma}$ as \cite{Lamarre1986}
\begin{equation}
    \text{NEP}_{\gamma}^2 \simeq 2 h \nu P_{\text{opt}} + \frac{2P_{\text{opt}}^2}{m\Delta\nu},
    \label{eqNEPphoton}
\end{equation}
where $h$ is Planck's constant, $\nu$ is the frequency of the radiation, $\Delta \nu$ is the spectral bandwidth and $m$ the number of polarizations detected, in the case of polarimetric measurements of the CMB, $m = 1$.

\section{The Magnetic Microbolometer (MMB)}

\begin{figure}
    \centering
    \includegraphics[width=0.8\columnwidth]{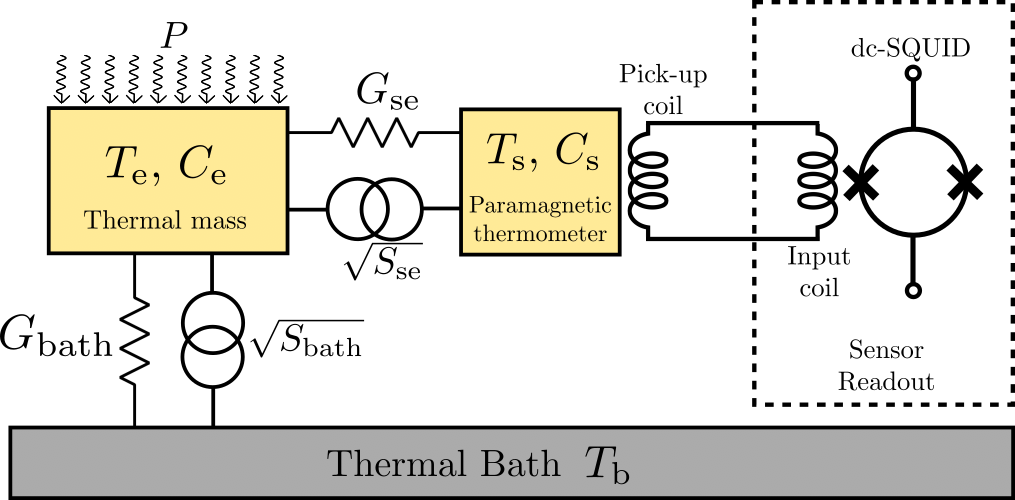}
    \caption{Schematics of the thermal model and readout scheme of an MMB. Incoming power $P$ is absorbed by a thermal mass with heat capacity $C_{\text{e}}$ that is therefore heated to a temperature $T_{\text{e}}$. This thermal mass is linked to a thermal bath with temperature $T_{\text{b}}$ by the weak thermal link $G_{\text{bath}}$. A paramagnetic temperature sensor with heat capacity $C_{\text{s}}$ is linked to the thermal mass by a second thermal link $G_{\text{se}}$ and is heated to a temperature $T_{\text{s}}$. The paramagnetic thermometer signal is transduced by a superconducting pickup coil magnetically coupled to a low-noise dc-SQUID. Additionally, the model includes the respective thermal noise sources related to the heat conductances.}
    \label{MMB_ThermalModel}
\end{figure}

Figure \ref{MMB_ThermalModel} shows the thermal model and readout scheme of the proposed MMB. Incoming radiation is absorbed and thermalized by a thermal mass in tight thermal contact to a paramagnetic temperature sensor that is biased with an externally applied weak magnetic field. The change in sensor magnetization, resulting from a temperature rise, induces an excess magnetic flux signal threading a superconducting pick-up coil that is coupled to a SQUID, a precise magnetometer capable of converting a magnetic flux signal into a voltage signal with quantum-limited accuracy \cite{Clarke2004}. The responsivity or gain, $\mathfrak{R}_{\text{MMB}}$, of the MMB is determined by the thermalization function $\partial T/\partial P$ that describes how variations in incoming power $P$ are converted into a variation in temperature $\delta T$ as well as the response function $\partial \Phi/\partial T$ of the paramagnetic thermometer that characterizes how the temperature variation $\delta T$ is transduced into a change of sensor magnetization $\delta M$ or magnetic flux variation $\delta \Phi$, respectively. In addition, the flux transfer function $\partial \Phi_\text{SQ}/\partial \Phi$ specifies the change of magnetic flux $\partial \Phi_\text{SQ}$ within the SQUID loop due to the change of magnetic flux $\partial \Phi$  threading the pickup coil has to be taken into account. Looking at the flux variation $\delta \Phi_{\text{SQ}}$ within the SQUID loop as the actual measurement signal for incoming power $\delta P$, the responsivity $\mathfrak{R}_{\text{MMB}}$ of the detector can be written as:
\begin{equation}
    \mathfrak{R}_{\text{MMB}}=\frac{\partial\Phi_{SQ}}{\partial P}=\frac{\partial \Phi_{SQ}}{\partial \Phi}\cdot\frac{\partial \Phi}{\partial T}\cdot\frac{\partial T}{\partial P}.
    \label{genericTerms}
\end{equation}

\begin{figure}
\centering
    \includegraphics[width=0.9\columnwidth]{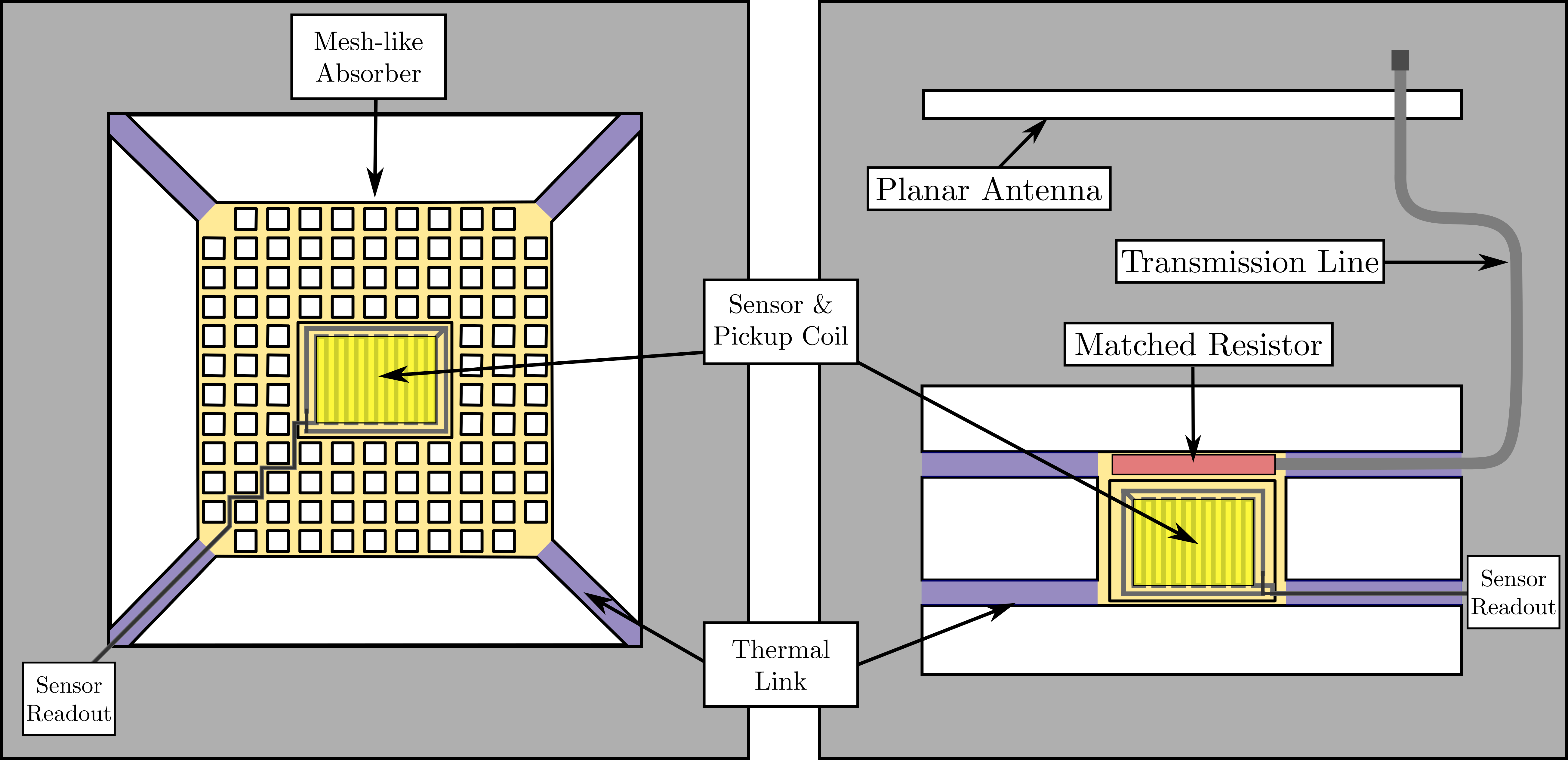}
    \caption{Conceptual detector layout of an MMB. Left: based on a thin metal mesh-like absorber with a surface resistivity tuned to the impedance of free space that absorbs incoming optical radiation. The subsequent rise in temperature is measured by a paramagnetic sensor and magnetically coupled pickup coil. Right: Antenna-coupled scheme, consisting of a planar antenna that couples radiation onto a microstrip transmission line terminated with a matched resistor thermally coupled to a paramagnetic material placed over a meander-shaped pickup coil which in both cases reads out the magnetization and simultaneously introduces the biasing magnetic field.}
  \label{MMB_Layout}
\end{figure}

To perform the measurement, incoming power in the form of millimeter-wave radiation, needs to be efficiently coupled to the detector. Radiation coupling can be based on either absorber-coupling \cite{Marnieros2020} or antenna-coupling schemes \cite{Kuo2008, Posada2018}. Sketches of each type of detector can be seen in Fig.\,~\ref{MMB_Layout}.

 In the following sections, we present a detailed discussion of the various factors in Eq.\,~\ref{genericTerms} applied to the proposed bolometer. 

\subsection{Paramagnetic Temperature Sensor}

In recent years, dilute alloys of gold (Au) or silver (Ag) as host materials that are doped with several hundred ppm of erbium (Er) has been established as state-of-the-art paramagnetic temperature sensors for MMCs \cite{Kempf2018}. An in-depth study of the thermodynamic properties of the resulting alloys Au:Er or Ag:Er can be found in Refs. \citeext{Enss2000} and \citeext{Fleischmann2005}. There isn't, however, extensive studies of the thermodynamic properties of these materials at higher temperatures. Simulations and measurements done in the past are mostly limited to temperatures below $100\,\text{mK}$ as MMCs are typically operated in that temperature range. For this reason, we developed a simulation framework to determine the thermodynamic properties of a Au:Er- or Ag:Er-based paramagnetic thermometer being operated in a bolometric application for CMB surveys.

The dilute alloys Au:Er or Ag:Er with several $100\,\mathrm{ppm}$ of erbium can be considered as a two-level system with effective spin $\Tilde{S}=1/2$ and g-factor $g = 6.8$ at temperatures below $1\,\text{K}$ \cite{Fleischmann2005}. 

A realistic thermodynamic model of Au:Er or Ag:Er can be obtained by considering the influence of the interactions between the magnetic moments in the lattice. Two mechanisms of spin-spin interaction are important to consider. The first is the magnetic dipole-dipole interaction, and the second is the so-called RKKY interaction. The latter is based on the indirect exchange interaction of the $\text{Er}^{\text{3+}}$ ions via the electrons in the conduction band of the host material. Here, an interaction parameter $\alpha$ is introduced that is the ratio between the strength of the RKKY and magnetic dipole-dipole interaction. $\alpha$ is a free parameter that is used to fit the simulation results to measured data as a final step.

To obtain the specific heat, $C_{\text{S}}(B,T)$, and magnetization, $M(B,T)$, accounting for the mentioned spin-spin interactions, a Monte-Carlo simulation was developed following the algorithms presented in Refs. \cite{Fleischmann2003} and \cite{Schonefeld2000}. A simulated crystalline fcc lattice is randomly populated with a given number of $\text{Er}^{\text{3+}}$ ions $N_{\text{s}}$ in a given volume of host material, the ratio of the $\text{Er}^{\text{3+}}$ ions to host atoms is adjusted by varying the simulated sample volume $V$. With the resulting arrangement, the Hamiltonian operator containing the Zeeman splitting function for a given externally applied magnetic field for each spin and the dipole-dipole and RKKY interaction operators between each pair of spins for all possible spin states is built. A diagonalization of the resulting Hamiltonian matrix is performed to extract the energy states of the current iteration. The entire process is repeated numerous times to account for various possible $\text{Er}^{\text{3+}}$ ion arrangements. With the results of this simulation, we can estimate the thermodynamic properties from the calculated energy states $E$ for each iteration as:
\begin{equation}
    C_{\text{S}} = \frac{N_{\text{s}}}{V k_{\text{B}}T^{2}}\Big{\{}\langle E^{2}\rangle - \langle E\rangle^{2}\Big{\}},
\end{equation}
\begin{equation}
    M = - \frac{N_{\text{s}}}{V}\Big{\langle} \frac{\partial E}{\partial B}\Big{\rangle},
\end{equation}
where $k_{\text{B}}$ is the Boltzmann constant and $T$ is the temperature.
Taking the partial derivative of the last expression with respect to temperature yields \cite{Fleischmann2003}:

\begin{equation}
    \frac{\partial M}{\partial T} = \frac{N_{\text{s}}}{V k_{\text{B}}T^{2}}\Big{\{}\Big{\langle} E \frac{\partial E}{\partial B}\Big{\rangle} - \langle E\rangle \Big{\langle} \frac{\partial E}{\partial B}\Big{\rangle}  \Big{\}}.
\end{equation}

\begin{figure}
    \centering
        \includegraphics[width=\columnwidth]{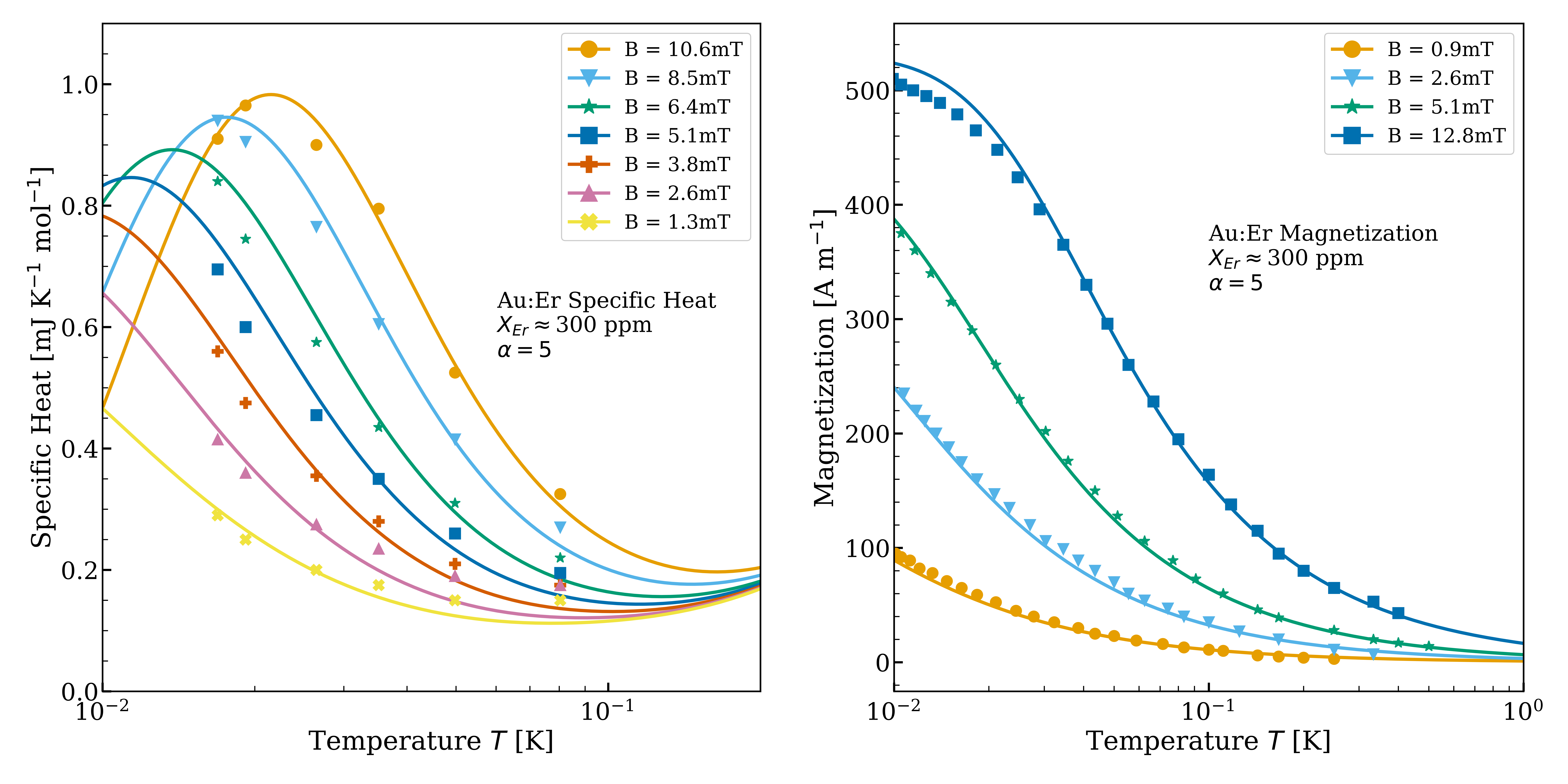}

        \caption{Results of our Monte-Carlo method based calculation of the specific heat (Left) and magnetization (Right) of a Au:Er sample with a concentration $x_{Er} \approx 300$ppm biased with various magnetic fields. Solid lines represent our simulation results the ratio between the strength of the RKKY and magnetic dipole-dipole interaction ($\alpha$) was adjusted to fit the data and it's final value is five, and the data points are taken from measurements presented in Ref. \cite{Fleischmann2005}.}
        \label{SimComparison}
\end{figure}

To verify the simulation of the paramagentic sensor, measurement data presented in Ref. \cite{Fleischmann2005} were used as reference. These measurements were extracted from a Au:Er sample with an Er concentration of $x_\mathrm{Er} \approx 300\,\mathrm{ppm}$. In Fig.\,\ref{SimComparison}, the measured data for the specific heat and magnetization of said sample are presented with superimposed simulated curves using our scripts. The specific heat shown also accounts for the free electron contribution. This electron component shows a linear dependence with temperature and can be estimated as $C_{\text{e}} = \gamma \cdot T$, where $\gamma = 6.9\times10^{-4} $ $\text{J}\text{mol}^{-1}\text{K}^{-2}$ is the Sommerfeld constant of gold \cite{Phillips1971}. From this comparison we can observe that measured data can be accurately estimated with the simulation setup described. It is worth mentioning that, recently, simulations results produced independently by another group were published in Ref.  \cite{Herbst2022} and that our results are in good agreement in the temperature range from $50\,\text{mK}$ to $1\,\text{K}$.

\begin{figure}

    \includegraphics[width=\columnwidth]{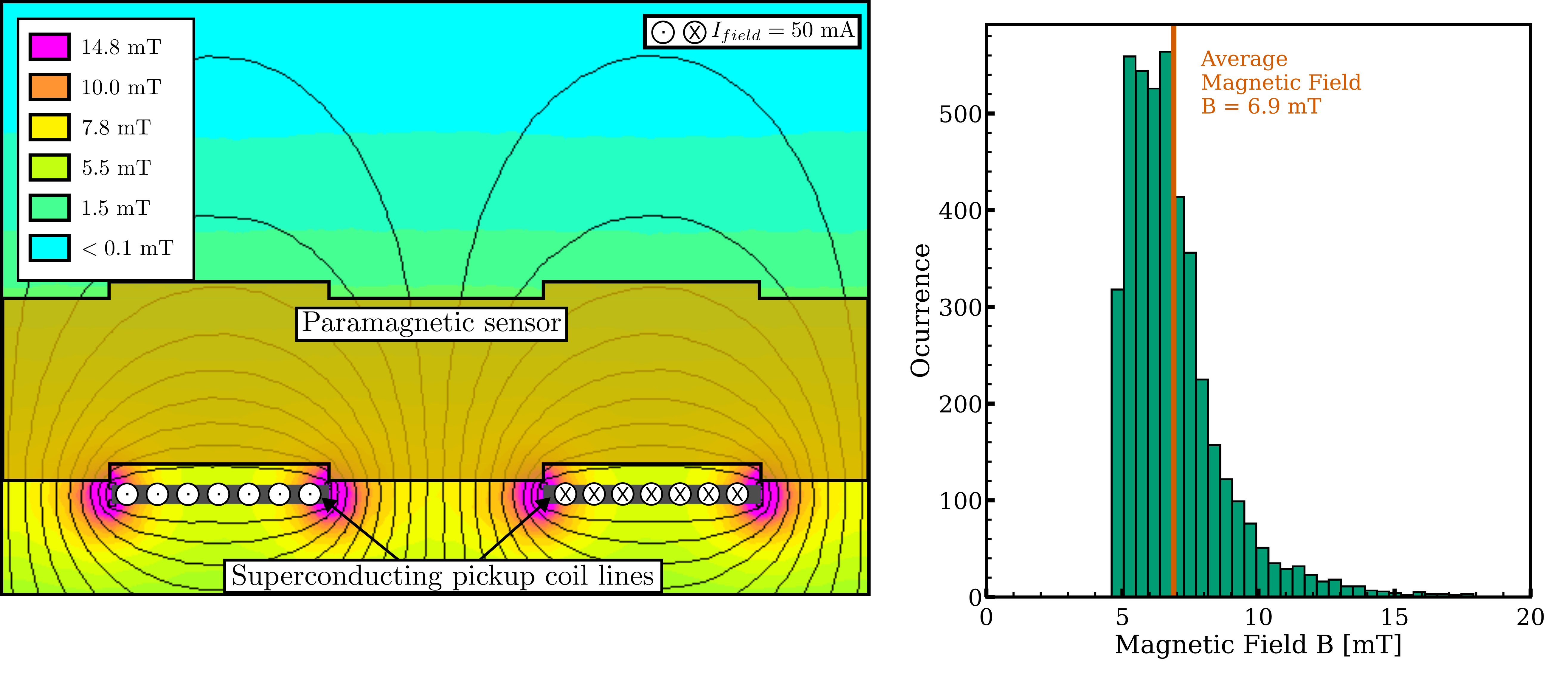}
    \caption{Calculated magnetic field distribution within a meander-shaped planar inductor through which a field current of $I_{\text{field}} = 50\,\text{mA}$ is running. The meander-shaped inductor is assumed to have a linewidth of $2.5\,\mu \text{m}$ and a pitch of $5\,\mu \text{m}$.}
    \label{fFEMMSim}
\end{figure}

To determine the temperature to magnetic flux conversion, a pickup coil geometry is defined and the magnetic field distribution due to a constant current is determined. In most state-of-the-art MMCs the geometry used is a meander-shaped planar inductor \cite{Kempf2018}. A persistent supercurrent, $I_{\text{field}}$, circulating through this coil magnetizes the sensor and simultaneously reads out flux variations due to flux conservation of closed superconducting circuits. FEMM was used to determine the constant magnetic field distribution of a meander-like pickup coil. From the obtained distribution, shown in Fig.\,\ref{fFEMMSim}, and subdividing the sensor in small cells, we can extract the derivative of the sensor's magnetization with respect to temperature for each cell using the results of the Monte-Carlo simulation. The temperature to flux conversion can then be determined as
\begin{equation}
    \frac{\partial \Phi}{\partial T} = \bigintss_V  \frac{|\vec{B}(\vec{r})|}{I_{\text{field}}}   \frac{\partial M}{\partial T}\Bigr|_{|\vec{B}(\vec{r})|} d^3r,
\end{equation}
integrating over the entire sensor volume $V$. The pitch and linewidth of the meander inductor have a strong influence in the results obtained. The smaller the pitch (and subsequently smaller linewidths) produces an increase of the generated average magnetic field in the sensor volume at a constant field current. For this reason, micrometer structured meander-like pickup coils are preferred.

\subsection{Thermalization Function}

Bolometers used in CMB or astronomical observations in general are often designed as leg-isolated detectors to achieve a weak thermal link to the heat bath, $G_{\text{bath}}$. Here, the detector structure is placed on a suspended membrane held in place by narrow thin beams minimizing phonon transport of heat. These thin narrow beams can be tailored precisely using standard microfabrication techniques, making the weak thermal link a predictable, controllable and stable input parameter once properly characterized.

MMBs can be modeled as two independent thermodynamic subsystems that are connected to each other. On the one hand, there is a conduction electron system that gives the material a base heat capacity following a linear dependence with temperature. And on the other hand, there is a spin system inherited from the $\text{Er}^{3+}$ ions immersed in the Au or Ag lattice that presents an additional heat capacity that varies with temperature and externally applied magnetic field. A suitable thermal model for the MMB should consider these two interacting systems involved. The link between both subsystems is modeled as an additional heat conductance, $G_{\text{se}}$, and it is derived from the relaxation time of electron-spin interactions.

\begin{figure}
    \centering
    \includegraphics[width=0.7\columnwidth]{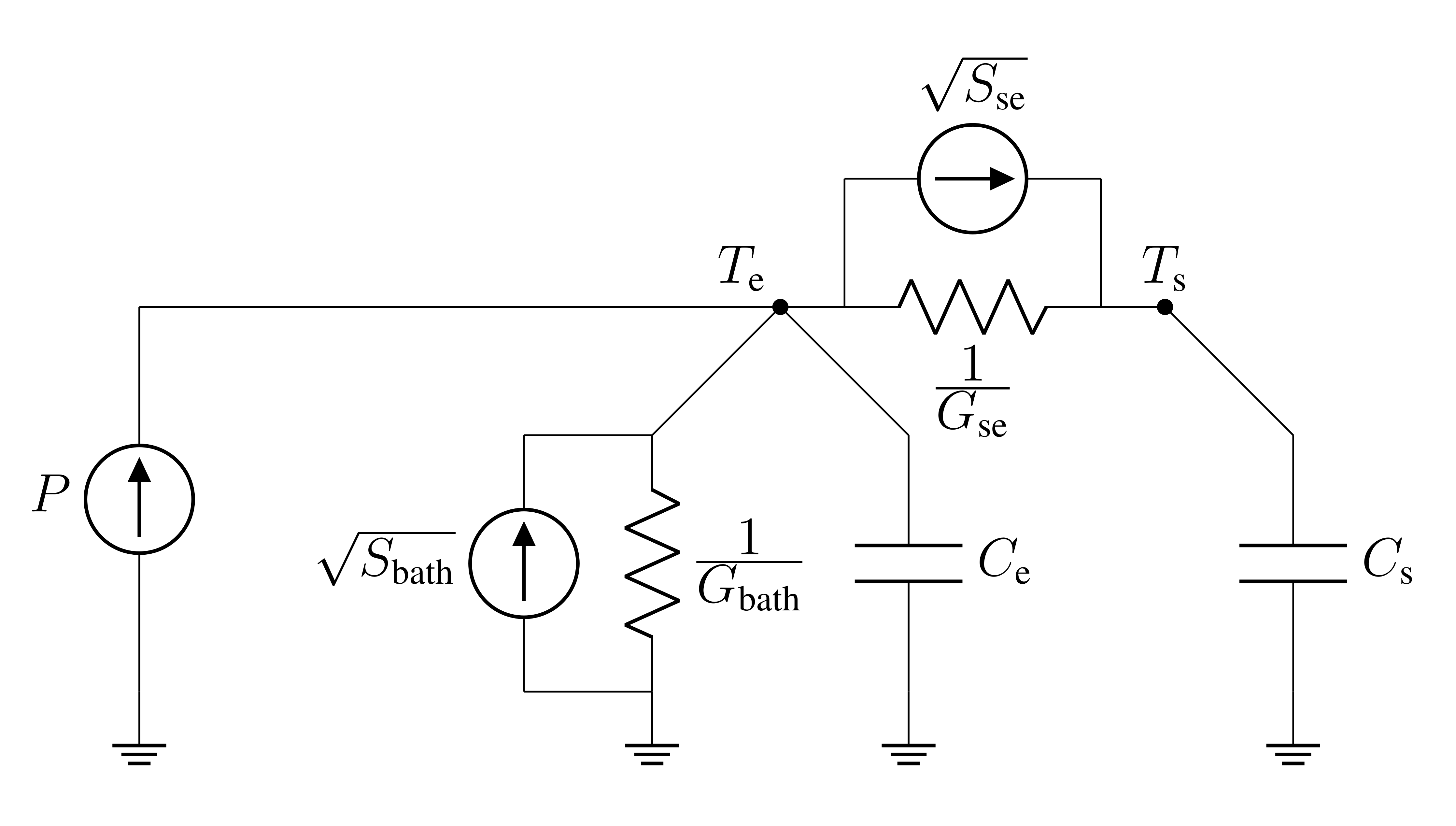}
    \caption{Equivalent electrical circuit model used to calculate the thermalization function $\partial T/\partial P$ of a magnetic microbolometer. Incoming radiation power $P$ heats the electron system to a temperature $T_{e}$. The spin system is heated to a temperature $T_s$ through the heat conductance $G_{se}$ that represents an effective thermal contact between sensor and absorber and accounts for the electron-spin relaxation time. The electron system is linked to a thermal bath through $G_{bath}$. $\sqrt{S_{bath}}$ and $\sqrt{S_{se}}$ are the phonon and spin thermal fluctuation noise contributions in the detector.}
    \label{thermalCircuit}
\end{figure}

A practical approach to solve the thermal model for these detectors is to use an equivalent electrical circuit, replacing heat capacities with capacitors and thermal conductance with resistors. Moreover, incident radiation power is modeled as a current source injecting a current into the detector and the temperature of the different detector subsystems are expressed as voltages at different circuit nodes. Finally, thermal noise contributions due to energy fluctuations among the different thermodynamic subsystems can be added to this circuit to account for thermal fluctuation noise \cite{Mather1982}. Using this approach for the thermal model presented in Fig.\,\ref{MMB_ThermalModel} yields the equivalent circuit model of a magnetic microbolometer as shown in Fig.~\ref{thermalCircuit}. 

Using the space-state variable method, the thermalization function of the bolometer can be derived from the differential equations that describe the equivalent circuit model, arriving to the following expression:
\begin{equation}
    \frac{\partial T}{\partial P} = \frac{1}{G_{\text{bath}}} \cdot \left[ {1-\frac{C_{\text{e}} C_{\text{s}}}{G_{\text{bath}}G_{\text{se}}}\cdot \omega^2 + \frac{C_{\text{s}}(G_{\text{bath}} + G_{\text{se}}) + C_{\text{e}} G_{\text{se}}}{G_{\text{bath}}G_{\text{se}}}\cdot i\omega}\right]^{-1}.
    \label{ThermFunc}
\end{equation}

Since the thermalization of the bolometer is the slowest process, it also determines the frequency response of the whole detector.

\subsection{Magnetic bias field and readout scheme}

The magnetization change of the paramagnetic sensor is measured as a change of magnetic flux threading a superconducting pick-up coil that simultaneously provides the required bias magnetic field by means of a persistent current running inside the pickup coil.
A persistent current generates a constant magnetic field free of noise with which the sensor is biased. Variations in magnetization of the sensor generate excess currents through the pickup coil due to flux conservation in closed superconducting circuits. The excess current is measured by the SQUID as a flux signal that can be calculated as:
\begin{equation}
\frac{\partial\Phi_{\text{SQ}}}{\partial\Phi} \approx \frac{k\sqrt{L_{\text{in}}L_{\text{s}}}}{L_{\text{m}}+L_{\text{stray}}+L_{\text{in}}},
\label{eqSQUID}
\end{equation}
where $L_{\text{m}}$ is the inductance of the pickup coil, $L_{\text{in}}$ is the SQUID's input coil inductance with which the excess current is measured. $L_{\text{s}}$ is the SQUID's loop inductance, $k$ is the coupling factor between $L_{\text{in}}$ and $L_{\text{s}}$. $L_{\text{stray}}$ is the parasitic inductance derived from the superconducting lines and wire bonds that connects the detector to the SQUID.

For the readout of a large cryogenic detector array, devices like the Microwave SQUID Multiplexer, that have already been implemented for MMC arrays in the past~\cite{Kempf2014}, can be used.

\subsection{Detector Noise Estimation}

To analyze ultimate performance of an MMB, responsivity and detector bandwidth alone are not enough. The detector noise contributions need to be estimated to decide whether it will be capable of performing BLIP of the sky. The main contributions that must be considered for the MMB are the TFN, magnetic Johnson noise (J), erbium $1/f$ excess (ER) noise and readout SQUID (SQ) noise. These contributions referred to the detector input give

\begin{equation}
    \shortstack{NEP}^2_{\text{det}} =  \frac{S_{\text{th}}}{\big|\frac{\partial T}{\partial P}\big|^2} + \frac{S_{J,\Phi}}{\big|\frac{\partial T}{\partial P}\big|^2\big|\frac{\partial \Phi}{\partial T}\big|^2} + \frac{S_{\Phi , Er}}{\big|\frac{\partial T}{\partial P}\big|^2\big|\frac{\partial \Phi}{\partial T}\big|^2} + \frac{S_{\Phi_s}}{\big|\frac{\partial T}{\partial P}\big|^2\big|\frac{\partial \Phi}{\partial T}\big|^2  \big|\frac{\partial \Phi_s}{\partial \Phi}\big|^2}.
\end{equation}
The terms $S_{th}$, $S_{J,\Phi}$, $S_{\Phi, Er}$ and $S_{\Phi_s}$ are the spectral density of the respective noise source and will be described in the following paragraphs.

There exists two terms within TFN, one represents the thermal fluctuation of phonon energy in the thermal link to the heat bath and the other, the energy fluctuations between the conduction electrons and the spin subsystems within the sensor. The phonon term can be estimated by \cite{Mather1982}

\begin{equation}
    S_{\text{bath}} = 4 \gamma k_B T^2 G_{\text{bath}},
    \label{eqNoiseGbath}
\end{equation}
where 
\begin{equation}
\gamma = \frac{\beta+1}{2\beta+1}\cdot\frac{1-(T_{\text{bath}}/T)^{2\beta+1}}{1-(T_{\text{bath}}/T)^{\beta+1}}
\end{equation}
is a numerical factor derived for materials at low temperatures and since there is only phonon transport of heat through the thermal link and no electron contribution, $\beta = 3$.

The energy fluctuations between the conduction electrons and the spin subsystems can be modeled and estimated by a similar method, arriving to the following expression \cite{Fleischmann2005}

\begin{equation}
    S_{\text{se}} = 4 k_B T^2 G_{\text{se}}.
\end{equation}

Therefore, the total TFN contribution to detector NEP is
\begin{equation}
    \frac{S_{\text{th}}}{\big|\frac{\partial T}{\partial P}\big|^2} = \frac{S_{bath} + S_{\text{se}}}{\big|\frac{\partial T}{\partial P}\big|^2} =  4 \gamma k_{\text{B}} T^2 G_{\text{bath}} + 4  k_{\text{B}} T^2 \frac{G_{\text{bath}}^2}{G_{\text{se}}}\cdot\bigg(1+\frac{(2\pi f)^2 C_{\text{e}}^2}{(G_{\text{bath}}+G_{\text{se}})^2}\bigg).
\end{equation}

Magnetic Johnson Noise of the MMB is produced by the thermally excited random movement of conduction electrons in the paramagnetic sensor, generating a fluctuating magnetic field that couples to the pickup coil seemingly as a part of the measured signal. This noise contribution can be calculated as \cite{Fleischmann2005}

\begin{equation}
    S_{\text{J},\Phi} = \mathfrak{K} \cdot \sigma \cdot k_\mathrm{B}T,
\end{equation}
where $\mathfrak{K}$ is a constant that depends on the geometry of the detector, $\sigma$ is the electrical conductance of the paramagnetic sensor and $T$ is the temperature. 

Each erbium ion in the sensor presents a fluctuating magnetic moment in the form of flicker noise, the origin of this magnetic fluctuation is not fully understood to date and is experimentally determined to be indpendent of temperature. Since it is a flicker noise it appears predominantly at low frequencies following a $1/f$ shape. It is commonly referred to as erbium $1/f$ excess noise. The spectral density of this contribution is considered to be:

\begin{equation}
    S_{\Phi , Er} = N_{Er} \frac{|B(\bold{r})|^2}{I_f^2}   S_m(f) ,
    \label{ErExcessNoise}
\end{equation}
where $N_{Er}$ is the number of erbium ions in the sensor and $S_m(f) $ stands for the spectral density noise power of the z-component of the magnetic moments of the erbium ions, $S_m(f) = 0.12\mu_B^2 \,/\, f^{\eta} $ where $\eta \approx 0.8 - 1$ \cite{Kempf2018}.

Regarding the SQUID noise, it can be described in two parts, at low frequencies there is a $1/f$ dependence, typical values at 1\,Hz is in the range of $\sqrt{S_{\Phi_s} }= 3 - 4 \, \mu \Phi_0 /\sqrt{\shortstack{Hz}}$ and over a few kHz, a frequency independent noise appears, and typical values are approximately $\sqrt{S_{\Phi_s} }= 0.2 - 0.4\, \mu \Phi_0 /\sqrt{\shortstack{Hz}}$, where $\Phi_0$ is the magnetic flux quantum \cite{Drung2011}.

\section{Detector Design Considerations}

The MMB detectors count with several parameters that can be optimized to achieve certain sensitivity and detector bandwidth. In the following paragraphs, a brief description of each parameter and its effect on detector performance will be discussed. We will focus first on a bolometer design similar to the one described in Ref. \cite{Suzuki2016} suited for an antenna-coupled bolometer in which a termination load and temperature sensor are placed on a thin suspended structure as shown in Fig.~\ref{MMB_Layout} (Right).

The area of the paramagnetic sensor is mostly determined by the required pickup coil inductance. From Eq.\,\ref{eqSQUID} it can be determined that in this situation the optimal inductance value for the pickup coil is to be equal to the input inductance of the readout SQUID. We are assuming an input inductance of $L_{\text{in}} = 2\, \text{nH}$, as is often found in MMC arrays, for the meander-like inductor with a pitch of $5\,\mu\text{m}$ and trackwidth of $2.5\,\mu\text{m}$, we determined using FastHenry3 that the required area to match the SQUID's input inductance is $180\times180\, \mu \text{m}^2$.

Once the area of the paramagnetic sensor and pickup coil is established, as a first optimization task, the thickness of the paramagnetic sensor fabricated using thin-film deposition technology is determined. Even though a thicker sensor means more field lines will thread through it, the total average magnetic field within the sensor at some point will start to decrease since most of the biasing field will be concentrated in the vicinity of the pickup coil and, in addition, the heat capacity of the detector will become larger. There is then an optimal sensor height with which the magnetic field is efficiently coupled to the thermometer without resulting in a large thermal mass that would make the detector too slow for certain applications.

\begin{figure}

	\includegraphics[width=\textwidth]{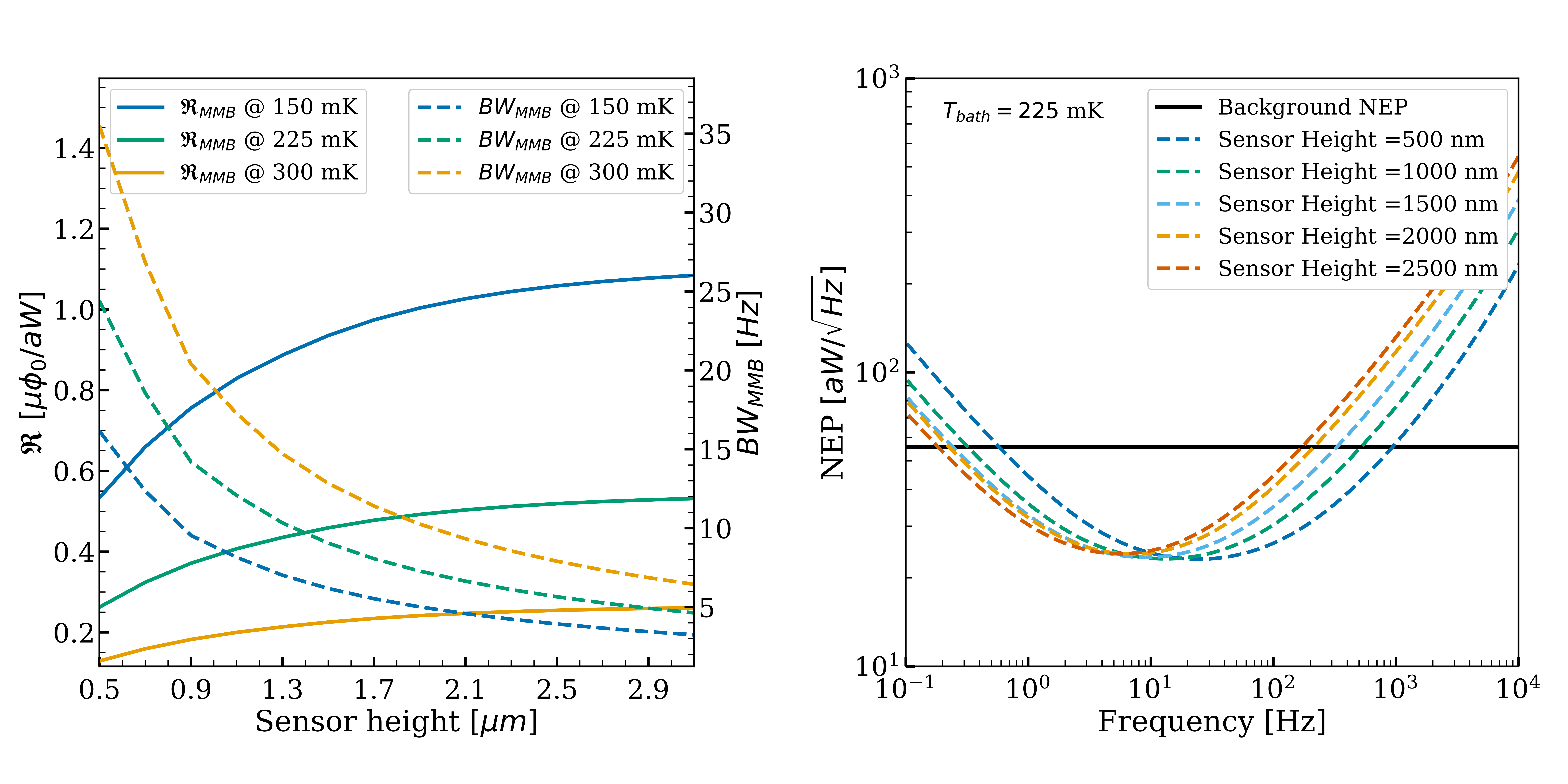}

	\caption{Left: $\mathfrak{R}_{\text{MMB}}$ and $BW_{\text{MMB}}$ of the MMB for different cryostat base temperatures as a function of sensor thickness. Right: colored dashed lines are the NEP$_{\text{det}}$ for multiple sensor heights at a cryostat base temperature of 225 mK, solid black line represents typical background NEP of $56~\text{aW}/\sqrt{\text{Hz}}$. A fixed leg length of $700\;\mu$m, $\text{Er}^{3+}$ concentration of 986 ppm and a field current of 100 mA was used.}
  \label{Resp_HSensor}
\end{figure}

In Fig.~\ref{Resp_HSensor} (Left) the Responsivity, $\mathfrak{R}_{\text{MMB}}$ and 3dB Bandwidth, $BW_{\text{MMB}}$, as a function of sensor height is shown for three different cryostat base temperatures that are commonly used in modern CMB instruments. We observe that $\mathfrak{R}_{\text{MMB}}$ increases with increasing sensor thickness until reaching a plateau in the range of $600 - 1500\,\text{nm}$, meaning that increasing the thickness even further would not noticeably improve sensitivity. The $BW_{\text{MMB}}$, on the other hand, decreases with increasing thickness due to a larger heat capacity. Regarding noise, in Fig.~\ref{Resp_HSensor} (Right) we can observe how the NEP$_{\text{det}}$ changes for different sensor heights at a cryostat base temperature of 225 mK. It can be observed how the NEP curves shift to the left according to the decreasing $BW_{\text{MMB}}$.

The leg length of the beams that hold the suspended structure in place for the design shown in Fig.~\ref{MMB_Layout} (Right) can be changed in a wide range of values, the main limitation is the mechanical stability of the finished device. This design parameter can be used to adjust the thermal conductance of the detector to the heat bath. We already know from Eq.\,\ref{eqRespBasic} and Eq.\,\ref{eqTimeBasic} that the smaller the thermal conductance to the heat bath, the larger the response of an ideal bolometer is and the slower it becomes. From the steady-state and frequency response expressions derived for the MMB one would expect a similar behavior. On the other hand, reducing the heat conductance will also modify the phonon thermal fluctuation noise and needs to be taken into account when adjusting this parameter.

\begin{figure}

	\includegraphics[width=\textwidth]{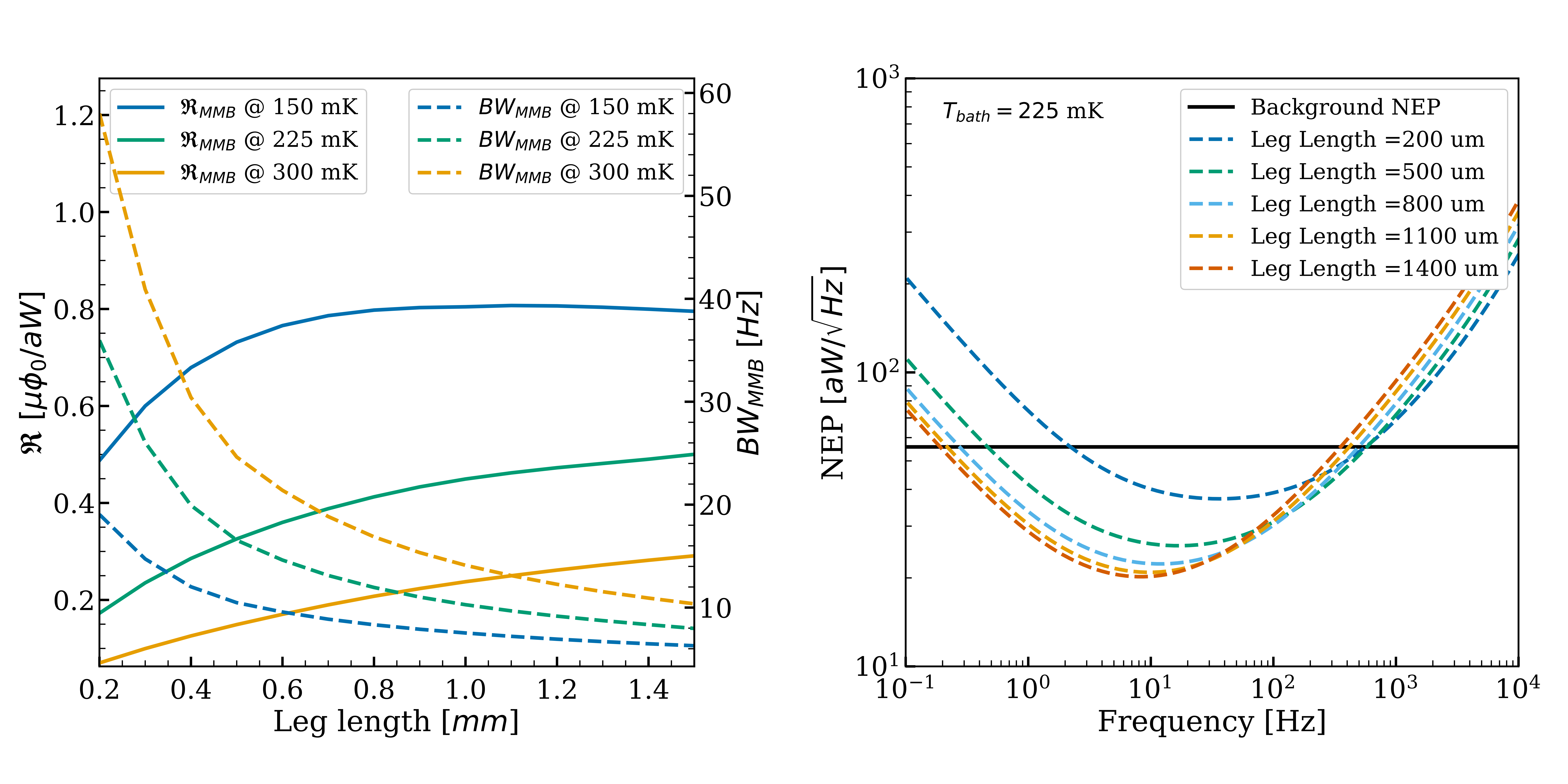}

	\caption{Left: $\mathfrak{R}_{\text{MMB}}$ and $BW_{\text{MMB}}$ of the MMB for different cryostat base temperatures as a function of the suspended structure's leg length. Right: colored dashed lines are the NEP$_{\text{det}}$ for multiple leg lengths at a cryostat base temperature of 225 mK, solid black line represents typical background NEP of $56~\text{aW}/\sqrt{\text{Hz}}$. A fixed sensor thickness of $1\;\mu$m, $\text{Er}^{3+}$ concentration of 986 ppm and a field current of 100 mA was used.}
   \label{Resp_LegL}
\end{figure}

In Fig. \ref{Resp_LegL} (Left) the $\mathfrak{R}_{\text{MMB}}$ and $BW_{\text{MMB}}$ is depicted as a function of leg length. It can be seen, that increasing the length of the suspending legs and thus reducing the heat conductance would not increase the responsivity indefinitely, this is because reducing the heat conductance makes the operating temperature of the MMB to increase due to the average incoming power measured. When the operating temperature increases, the magnetization variation of the paramagnetic sensor is reduced, and thus the response of the sensor is weaker. In Fig. \ref{Resp_LegL} (Right) we can observe how the NEP$_{\text{det}}$ changes with leg length at a cryostat bath temperature of 225 mK. For leg lengths that are above $800~\mu m$, the NEP$_{\text{det}}$ is not further reduced indicating an optimal value for this parameter.

\begin{figure}
	\includegraphics[width=\textwidth]{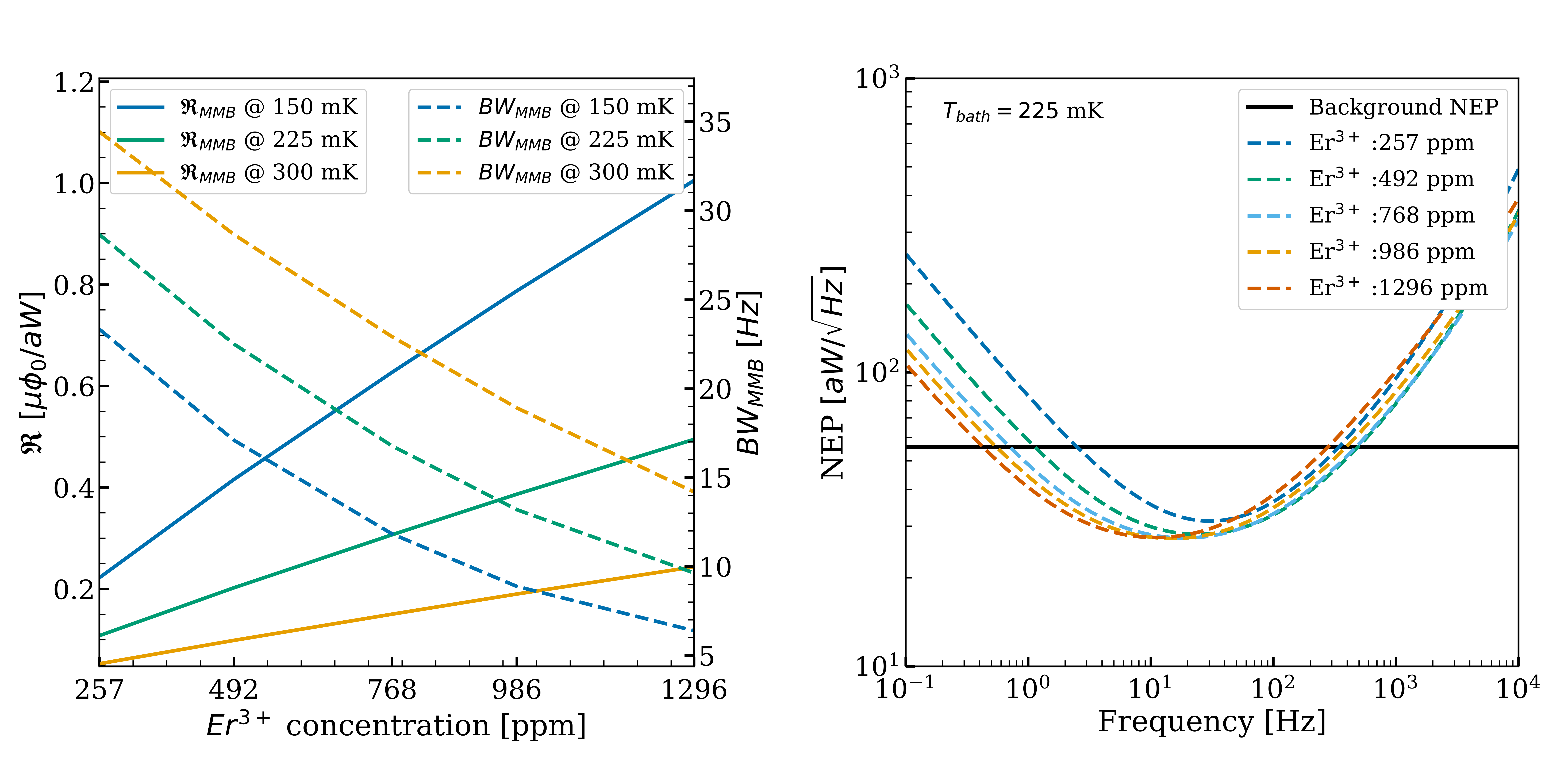}

	\caption{Left: $\mathfrak{R}_{\text{MMB}}$ and $BW_{\text{MMB}}$ of the MMB for different cryostat base temperatures as a function of $\text{Er}^{3+}$ concentration in the paramagnetic sensor. Right: colored dashed lines are the NEP$_{\text{det}}$ for multiple erbium concentrations at a cryostat base temperature of 225 mK, solid black line represents typical background NEP of $56~\text{aW}/\sqrt{\text{Hz}}$. A fixed leg length of $700\;\mu$m,sensor thickness of $1\;\mu$m and a field current of 100 mA was used.}
  \label{Resp_Cppm}
\end{figure}

Increasing the $\text{Er}^{3+}$ concentration of the paramagnetic material will enhance the signal response in an almost linear way, as seen in Fig.~\ref{Resp_Cppm} (Left). Although in principle, increasing the erbium concentration would always enhance sensitivity, it will also linearly increase the 1/f excess noise of the erbium ions as seen in Eq.\,\ref{ErExcessNoise}. In addition, the detector would be slower since it results in a higher specific heat of the spin subsystem. Fig.~\ref{Resp_Cppm} (Right) depicts how the NEP$_{\text{det}}$ is modified by changing the erbium concentration. The curves shift to the left due to the decreasing $BW_{\text{MMB}}$ and since the erbium excess noise increases, the curves start to meet towards the low frequency end of the spectrum for concentrations exceeding 1000 ppm.

\begin{figure}

	\includegraphics[width=\textwidth]{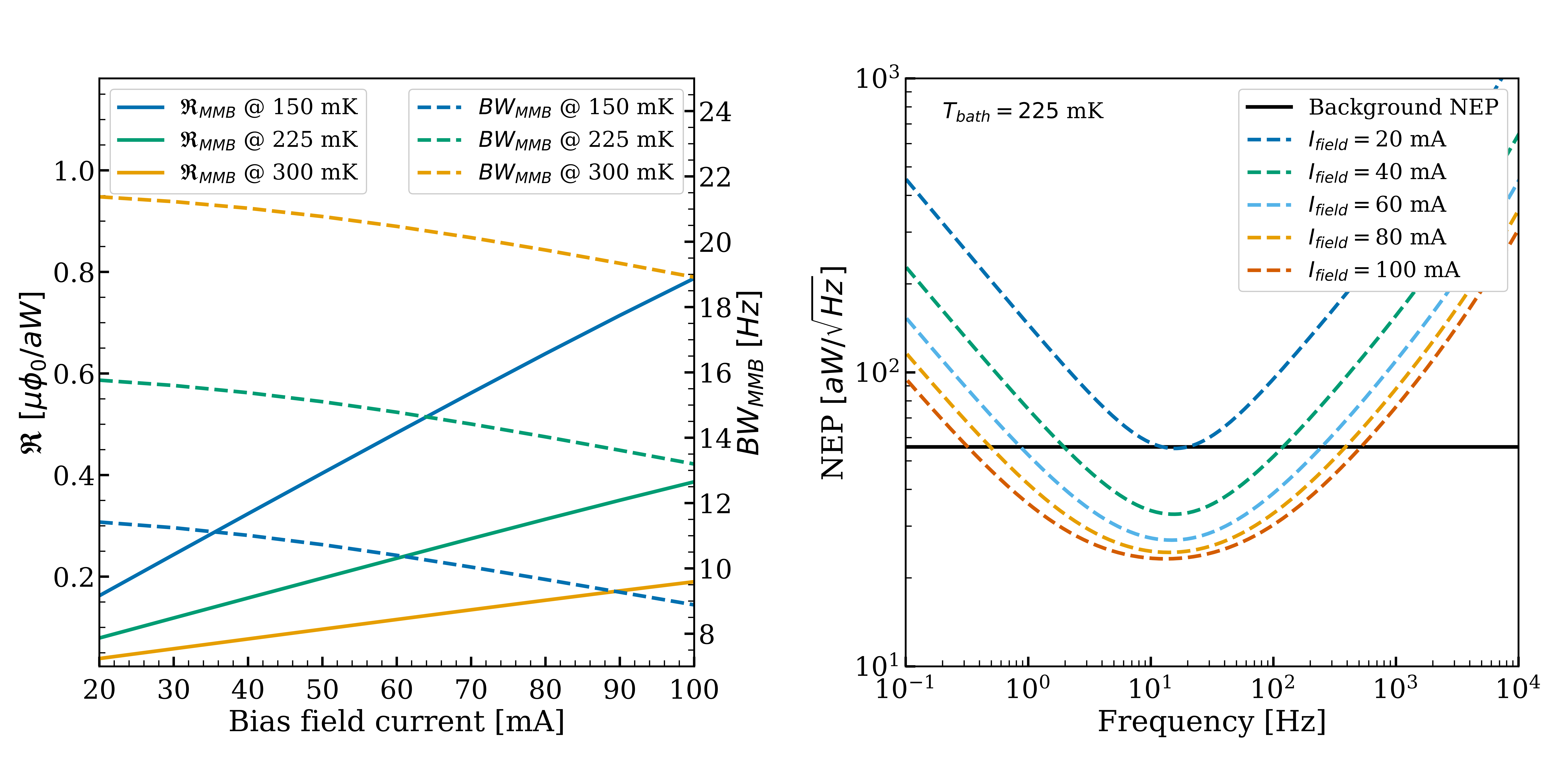}
    
	\caption{Left: $\mathfrak{R}_{\text{MMB}}$ and $BW_{\text{MMB}}$ of the MMB for different cryostat base temperatures as a function of the magnetic bias field current. Right: colored dashed lines are the NEP$_{\text{det}}$ for multiple bias field currents at a cryostat base temperature of 225 mK, solid black line represents typical background NEP of $56~\text{aW}/\sqrt{\text{Hz}}$. A fixed leg length of $700\;\mu$m,sensor thickness of $1\;\mu$m and an $\text{Er}^{3+}$ concentration of 986 ppm.}
  \label{Resp_IField}
\end{figure}

The persistent field current generates the magnetic field that magnetizes the paramagnetic thermometer. As seen in Fig.~\ref{Resp_IField} (Left) increasing this parameter enhances the signal response linearly without significantly affecting the $BW_{\text{MMB}}$. In Fig.~\ref{Resp_IField} (Right) it can be observed how NEP$_{\text{det}}$ is decreased with increasing $I_{\text{field}}$. This is mainly due to the fact that the responsivity is increased while maintaining the erbium excess noise contribution unmodified, therefore, at a certain point thermal fluctuation noise starts to predominate. This tells us that $I_{\text{field}}$ should be selected to be as large as possible. The limitation with this parameter is the $I_{\text{c}}$ of the used superconductor in the pickup coil circuit. This limitation may be overcome by using thicker films of superconducting material to fabricate the pickup coils and/or using superconducting materials of the highest critical current density available.

When working with MMB detectors designed as absorber-coupled bolometers, as shown in Fig.~\ref{MMB_Layout} (Left) and used, for example, in QUBIC~\cite{Marnieros2020}, there are some minor considerations that need to be foreseen. One is the inclusion of an additional heat capacity to the simulation accounting for the metal absorber that thermalizes the incoming radiation and will make the detector slower, and the other is a limitation on the maximal achievable leg length to maintain mechanical stability of the large suspended meshlike absorber with a high fabrication yield, ultimately restricting sensitivity. Metal materials used as absorbers in general are chosen to present a high residual resistivity to match the impedance of free space with thin- or ultra-thin-films and low specific heat to maintain a high enough detector bandwidth.

\section{Suitability of the MMB for CMB experiments}

With the theory described, and the simulation framework developed, we analyze one particular instrument in which this type of detector could be introduced and discuss whether the resulting detector achieves the required sensitivity and bandwidth for background limited detection using modern CMB experiments as reference.

For this case study, we will consider a CMB instrument optimized for large scale structures (like BICEP2 \cite{Ade2014} or QUBIC \cite{Battistelli2011b}) that observes the sky in the frequency band of $150\,\text{GHz}$ with a $25\%$ of bandwidth, in which the radiance peak of the CMB is found. We assume a typical aperture size of the instrument of $0.3\,\text{m}$, implying $\Theta_{\text{beam}} = 0.39~\text{degrees}$ at the central frequency. We take typical values for the scan speed and average incoming optical power and assumed them to be $\Dot{\Theta} = 1~\text{deg/sec}$ and $P_{\text{opt}}=6\,\text{pW}$, respectively. For this instrument, the maximum allowed time constant of the detector is then $\tau \leq 62\,\text{ms}$ (see Eq.~\ref{eqMaxTimeConstant}) corresponding to a required detector bandwidth of $f_{\text{BW}} \geq 2.57 \, \text{Hz}$ and the photon or background NEP can be estimated to be $\text{NEP}_{\gamma} = 55.79 \, \text{aW/}\sqrt{\text{Hz}}$ (see Eq.~\ref{eqNEPphoton}). This value of $\text{NEP}_{\gamma}$ is commonly reported in CMB instruments like QUBIC, POLARBEAR and BICEP2. The resulting detector should then have a $\text{NEP}_{\text{det}} < \text{NEP}_{\gamma}$ and a time constant $\tau \leq 62\,\text{ms}$. We will address both absorber-coupled and antenna-coupled schemes.

Due to their very small size as well as the strong desire to ultimately produce large detector arrays, MMBs will, similar to MMCs, be microfabricated using thin-film technology. This allows for changing detector parameters within a certain parameter range that is set by the available fabrication techniques. Using a numerical detector optimization methods as discussed for example in \cite{Kempf2018}, an optimal detector configuration for given experimental boundary conditions could in principle be found. Ultimately, we will use such a multi-variable and iterative process to find the optimal configuration for the given CMB instrument. However, for the case study discussed within this paper, we waive this optimization procedure and assume instead a fixed set of detector parameters that are reported and often used for fabricating state-of-the-art magnetic microcalorimeters \cite{Fleischmann2009, Kempf2018}. For this reason, we fix the choice of materials as well as the width, pitch and thickness of involved structures.

Niobium (Nb) is the superconductor of choice for these devices, silicon oxide ($\text{SiO}_{\text{x}}$) as insulator and silicon nitride ($\text{Si}_3{N}_{4}$) for the suspended structure. The paramagnetic sensor is a volume of Au:Er with an Er concentration of $\sim 1000\,\text{ppm}$ that will lie over a meander-like superconducting pick-up coil with a line-width of $2.5\,\mu\text{m}$ and pitch of $5\,\mu\text{m}$. An insulating $\text{SiO}_{\text{x}}$ layer of $300\,\text{nm}$ separates the sensor from the pick-up coil. The sensor itself would have an area of $180\times 180\,\mu\text{m}$ and a height of $1\,\mu\text{m}$. The magnetic bias field current selected is $100\,\text{mA}$ as used in the MMC detectors at GSI/FAI\cite{Pies2009}. We will consider a thermal bath temperature of $250\,\text{mK}$ corresponding to the base temperature of cryostats comonly used in CMB experiments. 

For the absorber-coupled bolometer we consider a palladium (Pd) mesh-structure that lies over a $\text{Si}_3{N}_{4}$ membrane suspended by four legs that are $200 \,\mu\text{m}$ in length and have a cross-section area of $30 \,\mu\text{m}^2$ resulting in $G_{\text{bath}} = 277.3 \text{pW/K}$. This absorber adds a heat capacity to the thermal model that is of $3\,\text{pJ/K}$ at the operating temperature \cite{Phillips1971}. The suspending legs, in the case of the antenna-coupled bolometer are selected to be $700 \, \mu\text{m}$ in length resulting in $G_{\text{bath}} = 162.8 \text{pW/K}$.

The bolometer, in this analysis, consists of a single pixel detector that is readout by a dc-SQUID. We will assume a SQUID with a self-inductance of $50\,\text{pH}$, an input inductance of $2\,\text{nH}$ and a coupling factor of $0.57$ \cite{Kempf2015}.

The time constants are calculated to be $10.6\,\text{ms}$ for the antenna-coupled MMB and $16.7\,\text{ms}$ for the absorber-coupled MMB. Both remain below the $\tau \leq 62\,\text{ms}$ requirement.

\begin{figure}
    \includegraphics[width=\textwidth]{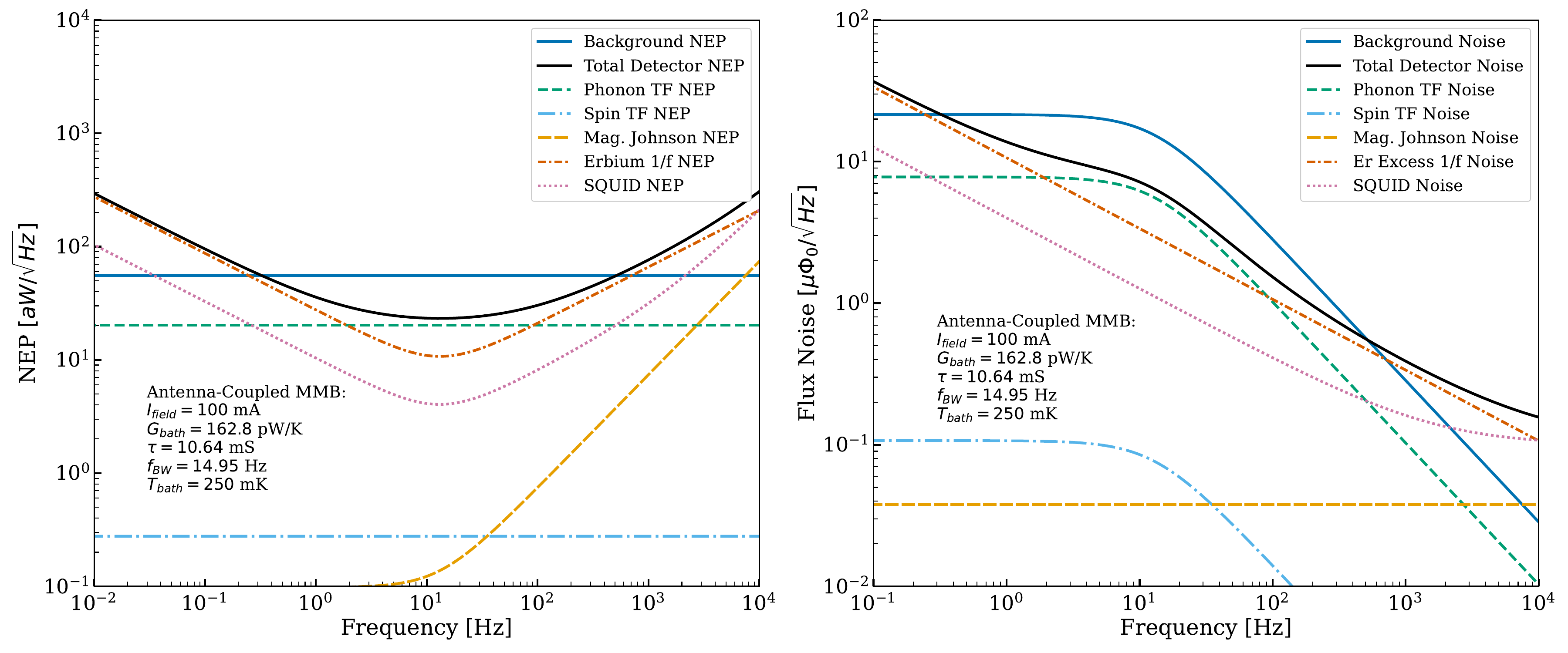}
    \caption{NEP (Left) and flux noise at the SQUID (Right) of the proposed MMB detector with antenna-coupled scheme. The solid blue line represents the background NEP/flux noise, the solid black line represents the total detector NEP/flux noise and dashed lines represent the detector's NEP/flux noise contributions discussed separately.}
    \label{NoiseContributions_antenna}
\end{figure}

\begin{figure}
    
    \includegraphics[width=\textwidth]{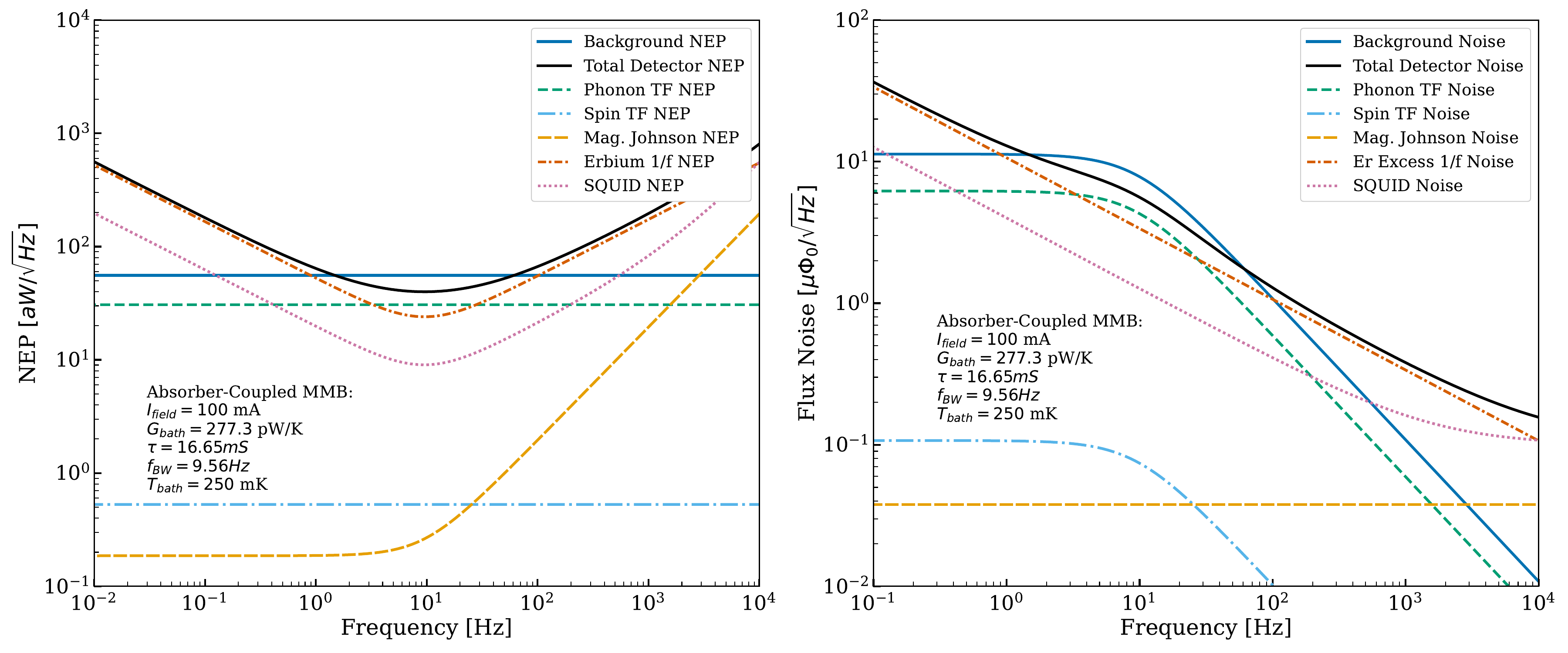}

    \caption{NEP (Left) and flux noise at the SQUID (Right) of the proposed MMB detector with absorber-coupled scheme. The solid blue line represents the background NEP/flux noise, the solid black line represents the total detector NEP/flux noise and dashed lines represent the detector's NEP/flux noise contributions discussed separately.}
    \label{NoiseContributions}
\end{figure}

In Fig. \ref{NoiseContributions_antenna} and \ref{NoiseContributions} the comparison between background noise and the different detector noise contributions in the case of the antenna-coupled and absorber-coupled MMBs respectively can be observed. All noises are shown both at the input of the detector (Left) as NEP and at the SQUID (Right) expressed as flux noise spectral density.

Ideally, phonon thermal fluctuation noise should be the second most predominant noise source, the first being background noise. If this condition is fulfilled, we know that the detector's performance is maximized and the proposed sensor technology reaches the ultimate sensitivity condition for the required operating temperature. In the two plots presented in Fig.\ref{NoiseContributions_antenna} and \ref{NoiseContributions} it can be seen that the total detector noise (solid black line) for a certain bandwidth lies underneath the background noise (solid blue line) this implies that the MMB presents background limited detection for said bandwidth. It is important to note that the unavoidable 1/f atmospheric noise is not included in these plots and therefore all 1/f noise contributions from the detector appear predominant towards the low frequency end of the spectrum. The main component of this low frequency noise is the erbium excess noise. To minimize the impact of this component, the sensitivity of the MMB should be enhanced to the point where background and thermal fluctuation noises dominate. There is one parameter that can increase sensitivity without significantly reducing detector bandwith or increasing the erbium excess noise contribution and this is the bias field current. Future optimization of MMBs would consist therefore in maximizing the I$_\text{c}$ of the used superconductor in the pickup circuit. This could be achieved either by increasing the thickness of the superconducting film up to some extent and/or utilizing superconducting materials with larger critical current densities J$_\text{c}$ than Nb.

\section{Conclusions}
In this paper we presented a magnetic microbolometer (MMB), a non-dissipative microstructured bolometer based on a paramagnetic thermometer composed of a normal metal material doped with erbium such as those found in MMC cryogenic detectors. A complete simulation framework was developed that accounts for the theoretical behavior of the resulting bolometer. A brief proof of concept case study was analyzed, taking into account typical constraints in CMB measurements and reproducible microfabrication techniques, to assess the suitability of metallic magnetic sensors in CMB experiments.

The results show that the detector proposed is a promising technology for CMB polarization measurements since their sensitivity can be optimized to realize background limited detection of the sky maintaining a low time response to avoid distortion of the point-source response of the instrument.

In addition to the possibility of background-limited detection we can enumerate certain advantages that MMBs have to offer. Due to their broad magnetization dependence with temperature, these detectors count with a very high dynamic range. The calibration of such a detector would be relatively simple as well and the absence of Joule dissipation simplifies the thermal design and means no Johnson noise is introduced in the readout. This sensor technology and its fabrication techniques are compatible with bolometric detector arrays already designed and installed in most CMB experiments and due to its dissipationless nature allows performing measurements at lower temperatures using the same cryostats already available in cosmology instruments.

It was observed that the predominant noise contribution at low frequencies for the particular MMB detector studied within this paper is 1/f excess noises associated with the erbium ions in the sensor. It's precise origin is yet unknown, however, the impact of this noise source can be overcome for example by increasing the bias field current, for this reason development and optimization of MMBs should be aimed at maximizing the critical current of the pickup circuit.


\bibliographystyle{spiejour}   
\bibliography{bib_tex}   



\end{spacing}

\end{document}